\documentclass[aps,pra,reprint,amsmath,amssymb,superscriptaddress,nobibnotes]{revtex4-1}
\usepackage{graphicx,SIunits}
\usepackage{bm}     
\usepackage{hyperref} 
\usepackage{dsfont}    
\usepackage{color}

\begin{document}


\title{Equitable multiparty quantum communication without a trusted third party}

\author{Tanumoy Pramanik}
\affiliation{Center for Quantum Information, Korea Institute of Science and Technology (KIST), Seoul, 02792, Republic of Korea}

\author{Dong-Hwa Lee}
\affiliation{Center for Quantum Information, Korea Institute of Science and Technology (KIST), Seoul, 02792, Republic of Korea}
\affiliation{Division of Nano \& Information Technology, KIST School, Korea University of Science and Technology, Seoul 02792, Republic of Korea}

\author{Young-Wook Cho}
\affiliation{Center for Quantum Information, Korea Institute of Science and Technology (KIST), Seoul, 02792, Republic of Korea}

\author{Hyang-Tag Lim}
\affiliation{Center for Quantum Information, Korea Institute of Science and Technology (KIST), Seoul, 02792, Republic of Korea}

\author{Sang-Wook Han}
\affiliation{Center for Quantum Information, Korea Institute of Science and Technology (KIST), Seoul, 02792, Republic of Korea}
\affiliation{Division of Nano \& Information Technology, KIST School, Korea University of Science and Technology, Seoul 02792, Republic of Korea}

\author{Hojoong Jung}
\affiliation{Center for Quantum Information, Korea Institute of Science and Technology (KIST), Seoul, 02792, Republic of Korea}

\author{Sung Moon}
\affiliation{Center for Quantum Information, Korea Institute of Science and Technology (KIST), Seoul, 02792, Republic of Korea}
\affiliation{Division of Nano \& Information Technology, KIST School, Korea University of Science and Technology, Seoul 02792, Republic of Korea}

\author{Kwang Jo Lee}
\affiliation{Department of Applied Physics, Kyung Hee University, Yongin, 17104, Republic of Korea}

\author{Yong-Su Kim}
\email{yong-su.kim@kist.re.kr}
\affiliation{Center for Quantum Information, Korea Institute of Science and Technology (KIST), Seoul, 02792, Republic of Korea}
\affiliation{Division of Nano \& Information Technology, KIST School, Korea University of Science and Technology, Seoul 02792, Republic of Korea}

\date{\today} 

\begin{abstract}
\noindent Multiparty quantum communication provides delightful applications including quantum cryptographic communication and quantum secret sharing. Measurement-Device-Independent (MDI) quantum communication based on the Greenberg-Horne-Zeilinger (GHZ) state measurement provides a practical way to implement multiparty quantum communication. With the standard {\it spatially localized} GHZ state measurement, however, information can be imbalanced among the communication parties that can cause significant problems in multiparty cryptographic communication. Here, we propose an equitable multiparty quantum communication where information balance among the communication parties is achieved without a trusted third party. Our scheme is based on the GHZ state measurement which is not spatially localized but implemented in a way that all the distant communication parties symmetrically participate. We also verify the feasibility of our scheme by presenting the proof-of-principle experimental demonstration of informationally balanced three-party quantum communication using weak coherent pulses.
\end{abstract}


\maketitle


{\it Introduction.--} Quantum key distribution (QKD) provides the information-theoretically secure way to share random bit strings between two remote parties~\cite{Bennett84,Ekert91}. Significant theoretical and experimental efforts have been dedicated to improve the security and practicality of QKD. For instance, Measurement-Device-Independent QKD (MDI-QKD), which is based on the entanglement detection in the middle of two communication parties, provides higher security than other ordinary QKD protocols since it is immune to all quantum hacking attempts to the measurement devices~\cite{braunstein12,Lo12,Choi16,park18}. Recently, MDI-QKD has been further improved to Twin-Field QKD (TF-QKD) that enables much longer communication distance~\cite{luca18,minder19,liu19,wang19,zhong19}. These remarkable works, however, are focussed on the secret communication between two parties. 

There are delightful multiparty quantum communication applications such as quantum cryptographic conferencing (QCC)~\cite{bose98,chen07}, and quantum secret sharing (QSS)~\cite{hillery99,cleve99,bell14}. These multiparty quantum communication protocols are usually based on distributing multipartite entanglement, e.g., the Greenberger-Horne-Zeilinger (GHZ) state. However, due to the difficulty of generating multipartite GHZ state, it is challenging to implement multiparty quantum communication. Indeed, there exists only a few proof-of-principle experiments of multiparty quantum communication~\cite{tittel01,chen05,gaertner07}, and long distance GHZ state distribution~\cite{erven14}. Remarkably, one can circumvent this difficulty by employing the MDI protocol based on the GHZ state measurement~\cite{fu15}. Despite the compromising performance, MDI quantum communication can be implemented using weak coherent pulses with decoy states~\cite{hwang03}. Therefore, it provides more practical solution for multiparty quantum communication than those based on the GHZ state generation. 

\begin{figure*}[t]
\includegraphics[scale=0.65]{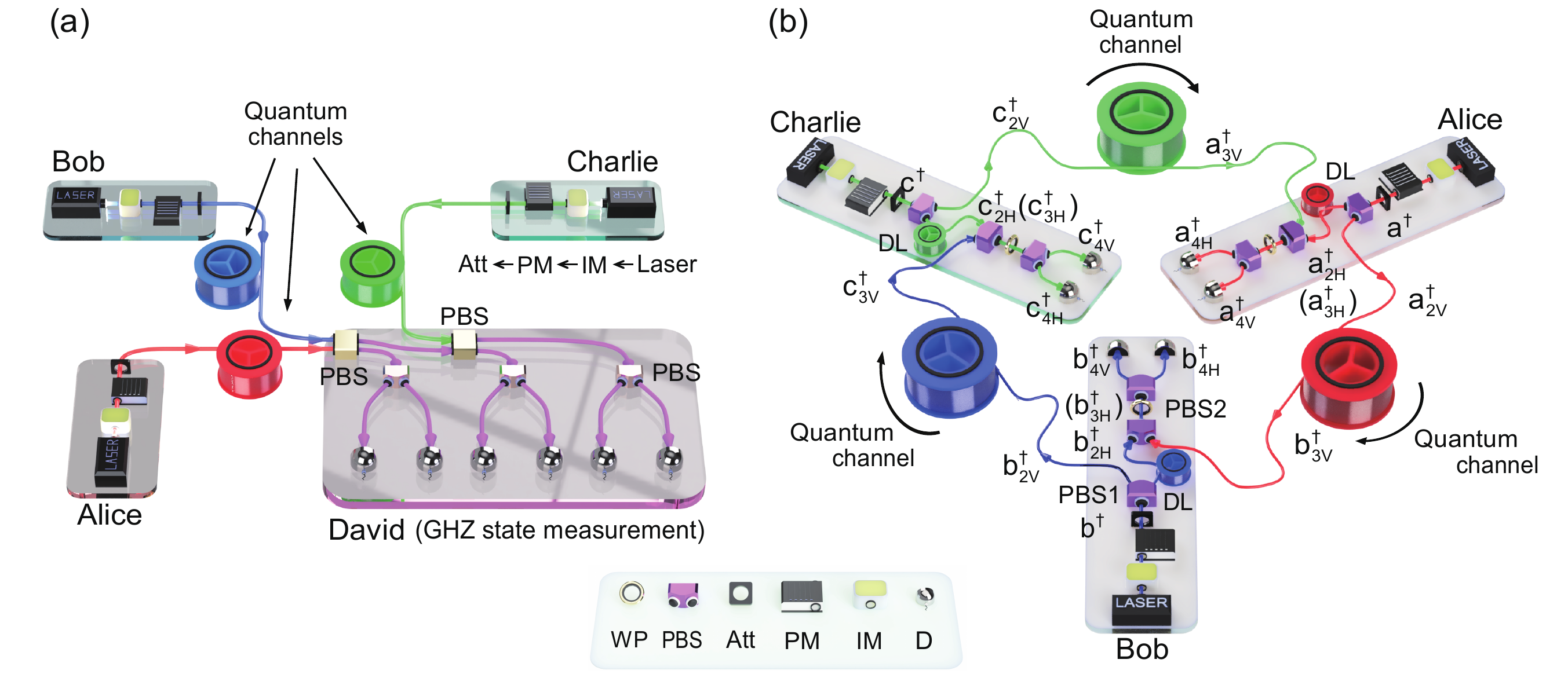} 
\caption{The schematics of (a) ordinary MDI multiparty quantum communication, and (b) equitable MDI multiparty quantum communication without a trusted third party. (a) In ordinary MDI multiparty quantum communication, the communication parties send optical pulses to David who performs the GHZ state measurement. (b) In the equitable MDI multiparty quantum communication, the GHZ state measurement is not performed {\it locally}, but all the communication parties equally participate in the GHZ state measurement. D : Single photon detector, IM : Intensity modulator, PM : Polarization modulator, Att. : Optical attenuator, PBS : Polarization beamsplitter, WP : Waveplate, DL : Delay line.}
\label{concept}
\end{figure*}

In order to fully enjoy the benefits of multiparty cryptographic communication, it is significant to retain the information equitability among the communication parties~\cite{shamir79,mao}. For instance, in secret sharing, each party has the same size of random bits $p_i$ as a private key and the secret is encoded as $s=p_1\oplus p_2\oplus\cdots\oplus p_N$. Here, $\oplus$ and the subscript $N$ denote the bitwise exclusive OR and the number of the communication parties, respectively. The secret $s$ can be restored only when all the $N$ parties cooperate. If, however, one knows others' private key, the secret can be reconstructed by less than $N$ parties. Therefore, for the security of secret sharing, the information equitability among the communication parties should be guaranteed. The information equitability should also be maintained when quantum communication is applied to multiparty cryptographic communication for the information-theoretically security.

In MDI multiparty quantum communication, it is assumed that the GHZ state measurement is performed by one of the communication parties or a third party~\cite{fu15}. This assumption is reasonable since entanglement detection cannot be performed via local operation and classical communication~\cite{horo08}. Note that the security of this scheme is based on the measurement induced entanglement, and thus, the third party can be considered as an eavesdropper. However, it causes information imbalance among the communication parties if one of them cooperates with the third party in secret of another communication parties. Considering the importance of information equitability in multiparty cryptographic communication, it is critical to rule out this information imbalance possibility.

In this Letter, we propose an equitable multiparty quantum communication protocol where information balance among the communication parties is achieved without a trusted third party. Our protocol is based on the GHZ state measurement which is not exclusively performed by one of the communication parties or a third party, but equally shared by all communication parties. We also show the feasibility of our protocol by presenting the proof-of-principle experimental demonstration of informationally balanced three-party quantum communication using weak coherent pulses.




{\it Standard MDI multiparty quantum communication.--} Let us briefly introduce MDI multiparty quantum communication protocol based on the GHZ state measurement~\cite{fu15}. Figure~\ref{concept}(a) shows a schematic diagram of conventional MDI three-party quantum communication. Each communication party (Alice, Bob, and Charlie) randomly assigns one of the four BB84 states to the optical pulse, then transmits it to a third party (David). The four BB84 states are the eigenstates of either $Z$-basis ($|0\rangle$ or $|1\rangle$) or $X$-basis ($|\pm\rangle=\frac{1}{\sqrt{2}}(|0\rangle\pm|1\rangle)$). By using polarization qubits, these states can be defined as $|0\rangle=|H\rangle$, $|1\rangle=|V\rangle$, $|\pm\rangle=|D/A\rangle=\frac{1}{\sqrt{2}}(|H\rangle\pm|V\rangle)$ where $|H\rangle$ and $|V\rangle$ denote horizontal and vertical polarization states, respectively. Then, David performs the GHZ state measurement, and announces the result, i.e., whether he obtained $|{\rm GHZ}^+\rangle$ or  $|{\rm GHZ}^-\rangle$, where $|{\rm GHZ}^\pm\rangle=\frac{1}{\sqrt{2}}(|000\rangle\pm|111\rangle)$. The standard linear optical setup to perform the three-photon GHZ state measurement is depicted in Fig.~\ref{concept}(a)~\cite{pan98}. Note that the standard GHZ state measurement is implemented in a {\it localized} region since the incoming photons should interact at polarizing beamsplitters (PBS). With the GHZ state measurement result, Alice, Bob and Charlie generate the sifted keys only when they have sent the optical pulses in the same basis. For the $Z$-basis state preparation, only when the input states were prepared as either $|000\rangle$ or $|111\rangle$, the GHZ state measurement outcomes (either $|{\rm GHZ}^+\rangle$ or $|{\rm GHZ}^-\rangle$) can be registered. Thus, each communication party can notice the others' bit information that is the same as her own. On the other hand, if the input states were prepared in $X$-bases, $|{\rm GHZ}^+\rangle$ ($|{\rm GHZ}^-\rangle$) is registered when the input states have a binary correlation of $X_A=X_B\oplus X_C$ ($X_A\oplus 1=X_B\oplus X_C$). Therefore, one can notice the binary correlation between the others' bits with the GHZ state measurement result and her own state. Note that the number of communication parties can be further increased since the GHZ state measurement scheme can be generalized with an arbitrary number of photons~\cite{pan98}. It is also remarkable that the MDI quantum communication can be implemented using weak coherent pulses with decoy states. The intrinsic quantum bit error rates (QBER) with the weak coherent pulses are $Q_Z=0$, and $Q_X=\frac{3}{8}$ for $Z$- and $X$-basis, respectively. We note that the non-zero QBER for $X$-basis can be mitigated by post-selecting the phase of weak coherent pulses~\cite{ma12, ma12b, fu15}.

In the following, we briefly present how the exclusive possession of the GHZ state measurement by a (third) party can affect to the information balance among the communication parties and the cryptographic communication. More complete analysis of information imbalance due to the exclusive possession of the GHZ state measurement can be found in Supplemental Materials. Let us assume that David who is supposed to perform the GHZ state measurement is under control of Alice, or he secretly cooperates with Alice. Consider the case where David does not perform the GHZ state measurement but perform simple projection measurement onto either all $X$ or $Z$ basis depending on the basis of Alice's state, and announce the faked GHZ state measurement result. For example, if Alice sends $|+\rangle$, David projects all the qubits onto the $X$ basis. Since the sifted key is obtained only when all three parties choose the same basis, David (or equivalently, Alice) can always notice Bob's and Chalie's bit information whenever a sifted key bit is constructed. However, Bob and Charlie cannot know the others' bit information. If Bob and Charlie want to check the QBER with this result, Alice can easily cheat them as if David has performed the GHZ state measurement correctly since she has all the bit information as well as the announced faked GHZ state measurement result. In other words, while Alice can obtain full information about Bob's and Charlie's secret keys, Bob and Charlie cannot notice her betrayal. Therefore, the exclusive possession of the GHZ state measurement provides information imbalance among the communication parties which is highly undesirable in multiparty cryptographic communication, e.g., quantum secret sharing~\cite{shamir79,mao}.



{\it Equitable multiparty quantum communication.--} The exclusive possession of the GHZ state measurement causes the information imbalance among the communication parties. Therefore, in order to have the information balance, the GHZ state measurement should be implemented in a way that all communication parties equally participate. The implementation of symmetrical GHZ state measurement among distant parties without a third party is not straightforward since entanglement cannot be increased by local operation and classical communication. 

In linear optical quantum information processing, two-qubit interaction can be implemented using two-photon interference and post-selection. It is remarkable that the origin of two-photon interference is not at the particle-particle interaction at a {\it localized} region, but interference between indistinguishable probability amplitudes~\cite{pittman96,kim13,kim14}. Recently, we have shown that two-photon entanglement generation and measurement can be implemented without the two photon overlapping at a localized region~\cite{kim18}. By applying this principle to multiple photon case, we can implement equitable multiparty quantum communication.

Figure~\ref{concept}(b) presents the schematic diagram of our proposal of the equitable multiparty quantum communication. Likewise the standard quantum communication, Alice, Bob, and Charlie prepare optical pulses in an eigenstate of either $X$- or $Z$-basis. Then, they send the probability amplitude of $|1\rangle$ to the next party, e.g., Alice $\rightarrow$ Bob, Bob $\rightarrow$ Charlie, and Charlie $\rightarrow$ Alice, while keeping that of $|0\rangle$. Then, they combine the receiving $|1\rangle$ state with their own $|0\rangle$ state and measure it in $X$-basis. The successful GHZ state measurement results are registered when each communication party receives a single-photon counting click, and thus a three-fold coincidence count is measured among Alice, Bob and Charlie. It is remarkable that the GHZ state measurement is not performed in a spatially localized region, but equally shared by all the communication parties of Alice, Bob, and Charlie. 

Let us describe how Fig.~\ref{concept}(b) performs the GHZ state measurement among distant parties. To prove the GHZ state measurement, let us consider the polarization GHZ input states of
\begin{eqnarray}
|{\rm GHZ}^{\pm}\rangle&=&\frac{1}{\sqrt{2}}\left(|HHH\rangle\pm|VVV\rangle\right)\nonumber\\
&=&\frac{1}{\sqrt{2}}\left(a_{1H}^{\dag}b_{1H}^{\dag}c_{1H}^{\dag}\pm a_{1V}^{\dag}b_{1V}^{\dag}c_{1V}^{\dag}\right)|0\rangle.
\end{eqnarray}
The operators $a^{\dag}$, $b^{\dag}$, and $c^{\dag}$ are the creation operators at Alice, Bob, and Charlie, respectively, and  the subscript $nP$ refers the spatial mode $n$ with the polarization state $P$. After the polarization beamsplitter (PBS1), the states evolve to
\begin{equation}
|{\rm GHZ}^{\pm}\rangle\rightarrow\frac{1}{\sqrt{2}}\left(a_{2H}^{\dag}b_{2H}^{\dag}c_{2H}^{\dag}\pm a_{2V}^{\dag}b_{2V}^{\dag}c_{2V}^{\dag}\right)|0\rangle.
\end{equation}
The probability amplitude exchange, i.e., keeping the probability amplitude $|0\rangle$ (equivalently, $|H\rangle$) while sending that of $|1\rangle$ (equivalently, $|V\rangle$) to the next party, is presented as 
\begin{eqnarray}
&&a_{2H}^{\dag}\rightarrow e^{i\theta_1}a_{3H}^{\dag},~~b_{2H}^{\dag}\rightarrow e^{i\theta_2}b_{3H}^{\dag},~~c_{2H}^{\dag}\rightarrow e^{i\theta_3}c_{3H}^{\dag},\label{transform1}\\
&&a_{2V}^{\dag}\rightarrow e^{i\phi_1}b_{3V}^{\dag},~~b_{2V}^{\dag}\rightarrow e^{i\phi_2}c_{3V}^{\dag},~~c_{2V}^{\dag}\rightarrow e^{i\phi_3}a_{3V}^{\dag}\label{transform2}
\end{eqnarray}
where $\theta_j$ and $\phi_j~(j=1,2,3)$ are the phase obtained during the probability amplitude exchange. Note that Eq.~\eqref{transform1} takes place via delay lines (DLs) which belong to the communication parties whereas Eq.~\eqref{transform2} happens via quantum channels (QCs) which can be accessed by eavesdroppers. After the PBS2 which combines two orthogonal polarization states into the same spatial mode, the states become
\begin{equation}
|{\rm GHZ}^{\pm}\rangle\rightarrow\frac{1}{\sqrt{2}}\left(a_{4H}^{\dag}b_{4H}^{\dag}c_{4H}^{\dag}\pm e^{i\Phi} a_{4V}^{\dag}b_{4V}^{\dag}c_{4V}^{\dag}\right)|0\rangle
\label{output}
\end{equation}
where $\Phi=\Sigma_j(\phi_j-\theta_j)$. Note that we can set $\Phi=0$ by adjusting the phase of the interferometer.

By transforming the polarization states using half waveplates (HWP) at $22.5^\circ$, the states become
\begin{eqnarray}
|{\rm GHZ}^+\rangle\rightarrow\frac{1}{2}&&\Big(a_{4H}^{\dag}b_{4H}^{\dag}c_{4H}^{\dag}+a_{4H}^{\dag}b_{4V}^{\dag}c_{4V}^{\dag}\nonumber\\
&&+a_{4V}^{\dag}b_{4H}^{\dag}c_{4V}^{\dag}+a_{4V}^{\dag}b_{4V}^{\dag}c_{4H}^{\dag}\Big)|0\rangle,\nonumber\\
|{\rm GHZ}^-\rangle\rightarrow\frac{1}{2}&&\Big(a_{4H}^{\dag}b_{4H}^{\dag}c_{4V}^{\dag}+a_{4H}^{\dag}b_{4V}^{\dag}c_{4H}^{\dag}\nonumber\\
&&+a_{4V}^{\dag}b_{4H}^{\dag}c_{4H}^{\dag}+a_{4V}^{\dag}b_{4V}^{\dag}c_{4V}^{\dag}\Big)|0\rangle
\end{eqnarray}
Thus, $|{\rm GHZ}^+\rangle$ state is registered as coincidences of $D_{HHH}$, $D_{HVV}$, $D_{VHV}$, and $D_{VVH}$, while $|{\rm GHZ}^-\rangle$ corresponds to the coincidences of $D_{HHV}$, $D_{HVH}$, $D_{VHH}$, and $D_{VVV}$, where $D_{ijk}$ denotes three-fold coincidences of $i$, $j$, and $k$ states at Alice, Bob, and Charlie. 

In order to verify the GHZ state measurement, it is essential that the other basis states than the GHZ states do not provide the same coincidence results. For three qubit states, there are six other basis states of $|HHV\rangle$, $|HVH\rangle$, $|HVV\rangle$, $|VHH\rangle$, $|VHV\rangle$, and $|VVH\rangle$. It is remarkable that these states does not provide three-fold coincidence among Alice, Bob, and Charlie due to the probability exchange transformation, Eqs.~\eqref{transform1} and \eqref{transform2}. Therefore, the proposed scheme successfully performs the GHZ state measurement. Note that this GHZ state measurement scheme can be implemented with an arbitrary number of qubits, so the equitable multiparty quantum communication can be implemented with an arbitrary number of communication parties.

The schematic of the present multiparty quantum communication scheme is inherently symmetrical without a third party. All the communication parties keep the probability amplitude of $|0\rangle$ and send that of $|1\rangle$ to the next party. The successful GHZ state measurement result is registered only when all the communication parties have photon counting clicks, therefore, all of them should cooperate in order to find out the GHZ state measurement result. These features provide the information balance among communication parties without introducing a third party. 

Although our protocol is based on the GHZ state measurement, unlike the MDI multiparty quantum communication~\cite{fu15}, it does not provide the MDI feature. Therefore, similar to the ordinary QKD protocols such as BB84 protocol, the whole communication party, including the measurement setup, should be protected from quantum hacking attempts by, for instance, monitoring the radiation towards the communication party~\cite{lydersen10, sajeed16}. We note that, in the MDI multiparty quantum communication, the measurement devices do not need to be protected, but the transmitters should be protected.

It is noteworthy that QCs, which eavesdroppers can access freely in quantum communication scenario, deliver very limited information, i.e., they only deliver $|V\rangle$ all the time. Therefore, it is undesirable for eavesdroppers to tap meaningful information via quantum channels without knowing the whole interferometer including DLs which is inside of the communication parties. Note also that Eqs.~\eqref{transform1} and \eqref{transform2} present that the individual transformations have no losses, or equivalently, have the identical losses. We remark that the loss imbalance between individual transformations can be balanced without introducing additional loss. Moreover, the loss analysis shows that any attempts altering the quantum channel loss increases the QBER in $X$-basis. We further discussed the potential issues on the practical implementation of our protocol, such as imbalanced channel losses and stabilizing polarization and phase, and proposed a practical scheme as well to address the practical issues, see Supplemental Materials for details.



\begin{figure}[b]
\includegraphics[width=3.5in]{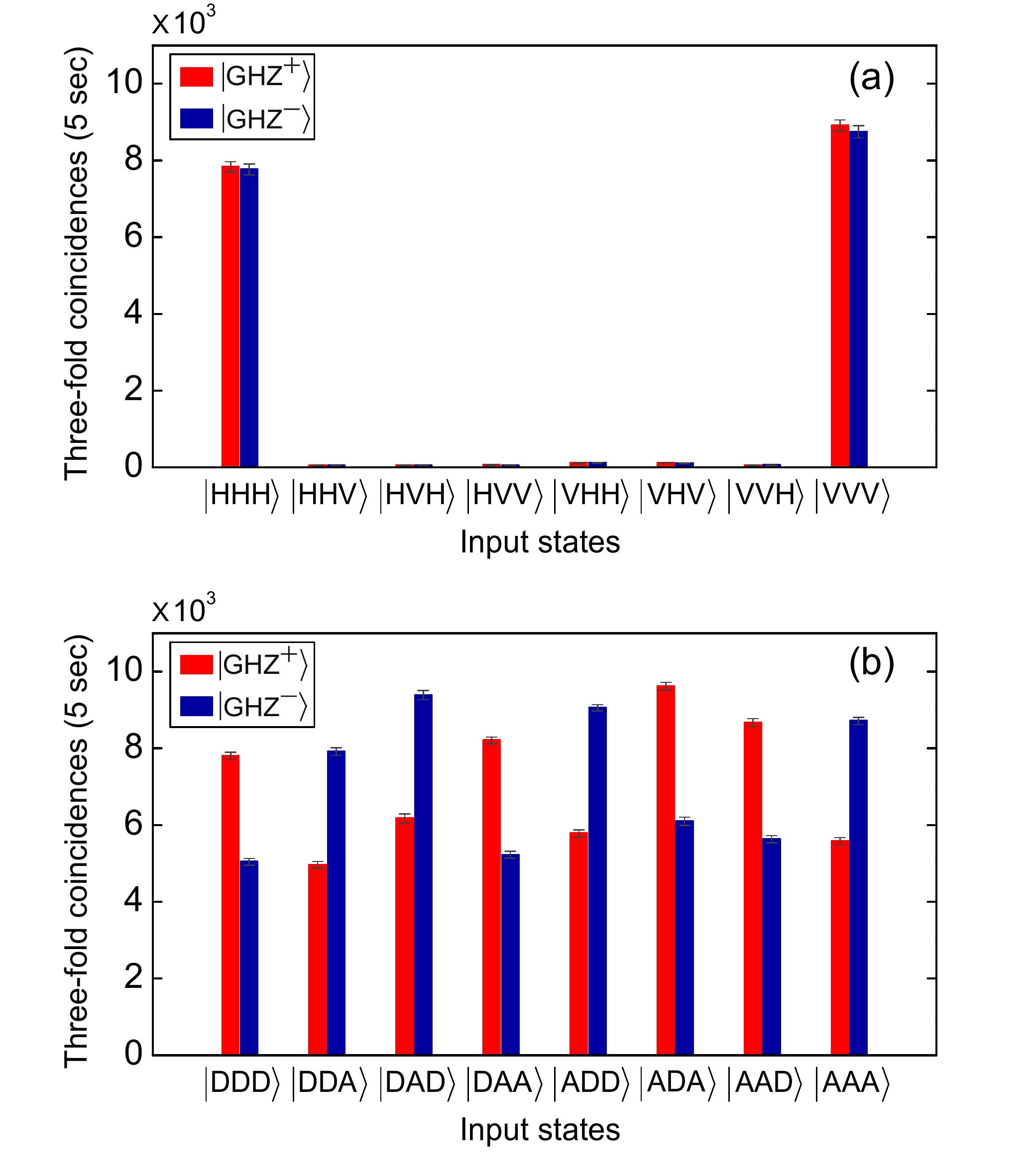} 
\caption{The GHZ state measurement results with various input states of (a) Z-basis and (b) X-basis. For Z-basis inputs, only $|HHH\rangle$ and $|VVV\rangle$ inputs provide the GHZ state measurement outputs. On the other hand, all the X-basis inputs give biased $|{\rm GHZ}^{\pm}\rangle$. The experimentally obtained QBER for Z- and X-basis are $Q_Z=2.56\pm0.32$ and $Q_X=39.1\pm0.34~\%$, respectively.} 
\label{data2}
\end{figure}

{\it Experimental results.--} The equitable multiparty quantum communication scheme of Fig.~\ref{concept}(b) requires synchronization and phase stabilization between photons traveling through different optical paths. We note that these technical demands has been realized with the current technology even when the optical paths are few hundreds of km in order to implement TF-QKD~\cite{liu19,wang19}, see also Supplemental Materials. Here, in order to verify the feasibility and practicality of the equitable multiparty quantum communication, we present the proof-of-principle experiment of three-party quantum communication on the optical table using the polarization state of weak coherent pulses. For details of the experiment, see Supplemental Materials.

We summarize the GHZ state measurement results with respect to various input states in Fig.~\ref{data2} (a) and (b) for $Z$-basis and $X$-basis inputs, respectively. During the data acquisition, the phase $\Phi$ was kept for $\Phi=0$. For the $Z$-basis inputs, only $|HHH\rangle$ and $|VVV\rangle$ inputs are registered as $|{\rm GHZ}^{\pm}\rangle$ while all other input states do not provide the three-fold coincidences. The QBER, $Q_Z=1-\frac{N_{HHH}+N_{VVV}}{\sum_{i,j,k=H,V}N_{ijk}}$ where $N_{ijk}$ denotes GHZ measurement output counts with $|ijk\rangle$ input, is measured as $Q_Z=2.56\pm0.32~\%$. Figure~\ref{data2}(b) shows the GHZ state measurement results with $X$-basis input states. As theoretically expected, the inputs of $|DDD\rangle$, $|DAA\rangle$, $|ADA\rangle$, and $|AAD\rangle$ provide the biased result towards $|{\rm GHZ}^+\rangle$, while the other states are biased to $|{\rm GHZ}^-\rangle$. The non-zero erroneous measurement results come from using weak coherent pulses instead of single photon inputs. The average QBER in $X$-basis is $Q_X=39.1\pm0.34~\%$. Here, the QBER is calculated as $Q_X=\frac{N_-}{N_+ + N_-}$ for $|DDD\rangle$, $|DAA\rangle$, $|ADA\rangle$, and $|AAD\rangle$ inputs and $Q_X=\frac{N_+}{N_+ + N_-}$ for the other inputs where $N_{\pm}$ refers the number of $|{\rm GHZ}^{\pm}\rangle$ measurement outcomes. Note that the experimentally obtained $Q_X$ is only about $2~\%$ higher than its theoretical value of $Q_X=37.5~\%$ with weak coherent pulse inputs~\cite{fu15}.


{\it Conclusion.--} In multiparty cryptographic communication, all the communication parties should have the same amount of information. Similarly, the information balance is essential for multiparty quantum communication. Here, we showed that the exclusive possession of GHZ state measurement in MDI multiparty quantum communication by one party causes information imbalance among the communication parties. In order to guarantee the information balance in multiparty quantum communication, we proposed equitable multiparty quantum communication based on the GHZ state measurement which is symmetrically shared by all the communication parties. We also verified the feasibility and practicality of the scheme by performing the proof-of-principle experiment using weak coherent pulses. Considering the importance of information balance in multiparty cryptographic communication, the equitable multiparty quantum communication will pave a new way towards secure communication.

\section*{Acknowledgement}
The authors thank Y. Shin for experimental assistance. This work is supported by the National Research Foundation of Korea (Grants No. 2019M3E4A1079777, No. 2019R1A2C2006381, and No. 2019M3E4A107866011), MSIT/IITP (Grants No. 2020-0-00972, and 2020-0-00947), and a KIST research program (No. 2E30620).

\end{document}